\documentclass[twocolumn,prl,aps,amsmath,amssymb]{revtex4}

\usepackage{graphicx}
\usepackage{dcolumn}
\usepackage{bm}

\usepackage{feynmp}

\begin{document}
\title{Critical Behavior of Non Order-Parameter Fields}
\author{\'Agnes {\sc M\'ocsy}}
 \email{mocsy@alf.nbi.dk}
 \author{Francesco {\sc Sannino}}
 \email{francesco.sannino@nbi.dk}
 \author{Kimmo {\sc Tuominen}}\email{tuominen@nordita.dk}
 \affiliation{NORDITA \& The Niels Bohr Institute, DK-2100 Copenhagen \O,
Denmark }
\date{January 2002}

\begin{abstract}
We show that all of the relevant features of a phase transition can be
determined using a non order parameter field which is a physical state
of the theory. This fact allows us to understand the deconfining
transition of the pure Yang-Mills theory via the physical excitations
rather than using the Polyakov loop.
\end{abstract}

\maketitle

\section{Introduction}
\label{uno}

Some of the relevant intrinsic properties of the phase transition
are best investigated using the order parameter of the theory.
Sometimes the order parameter can be a non physical object or not
experimentally accessible. The pure Yang-Mills deconfinement phase
transition constitutes a time honored example. Here the order
parameter is the Polyakov loop which is not a physical state of
the theory while the physical field is the glueball field. We
argue that it is always possible to extract all of the relevant
information about a specific phase transition using a non order
parameter field which is a physical state of the underlying
theory.

We predict the universal behavior of this field near and at the
phase transition. More specifically we predict a finite drop in
the mass of the singlet field at the phase transition. Our
predictions seem to reproduce surprisingly well the general
features of some lattice data and can be put to further and more
stringent tests via new first principle lattice simulations.
Finally, we define new physical quantities able to characterize
the phase transition.

\section{A renormalizable theory and critical behavior}
We consider a temperature regime close to the phase transition. In
order for our results to be as universal as possible we study a
renormalizable Lagrangian containing a field neutral under the
global symmetries and the order parameter field of theory as well
as their interactions. The protagonists of our theory are two
canonically normalized fields $h$ and $\chi$. The field $h$ is a
scalar of $Z_N$ while $\chi$ transforms according to
$\chi\rightarrow z\,\chi$ with $z \in Z_N$.

Here we consider the case $N=2$. This theory is suitable for
understanding the deconfining phase transition of 2 color
Yang-Mills which has been heavily studied via lattice simulations
\cite{Damgaard,{Hands:2001jn}}. A general, renormalizable,
potential is:
\begin{eqnarray}
V(h,\chi)&=&\frac{m^2}{2} h^2   + \frac{m^2_{\chi}}{2}\,
\chi^2 + \frac{\lambda}{4!}\chi^4 + g_{0}\,h \nonumber \\&+&
\frac{g_{1}}{2}\,h\chi^2 + \frac{g_2}{4}\,h^2\chi^2 +
\frac{g_3}{3!}h^3 + \frac{g_4}{4!}h^4 \ .
\end{eqnarray}
The coefficients are all real with $g_0=0$. Stability requires
$\lambda \geq 0$ and $g_4 \geq 0$. It is straightforward to
generalize this potential to $Z_N$. Since we will relate $\chi$ to
the Polyakov loop, we take it to depend only on space coordinates.
This leads to a mixed situation in which the physical states
depend also on time. We will also briefly comment on what happens
when the order parameter is taken to be four dimensional. The
Lagrangian we use to define our Feynman rules is:
\begin{eqnarray}
{\cal L}=\frac{1}{2}\partial_{\mu}h \partial^{\mu}h
-\frac{1}{2}{\nabla} \chi \cdot {\nabla} \chi -V(h,\chi) \ .
\end{eqnarray}
We start our investigation at temperatures below $T_c$. Since the
phase transition is of second order, $m_{\chi}$ vanishes at the
transition point. All of the coupling constants are such that $h$
and $\chi$ have zero vacuum expectation value in this regime.

We focus our attention on the $h$ self energy at the
one loop level at a given non zero temperature:
\begin{fmffile}{fmftempl}
\begin{eqnarray}
\parbox{15mm}{\begin{fmfgraph}(40,20) \fmfpen{thin}
  \fmfleft{i}\fmfright{o} \fmf{plain}{i,v}\fmf{plain}{v,o}
  \fmfv{decor.shape=circle, decor.filled=30, decor.size=.5w,
  label=$1PI$}{v}
\end{fmfgraph}} &=&12
\,\parbox{15mm}{
\begin{fmfgraph}(40,20)
  \fmfpen{thin}
  \fmfleft{i}\fmfright{o}
  \fmf{plain}{i,v,v,o}
\end{fmfgraph}} +2\,
\parbox{15mm}{
\begin{fmfgraph}(40,20)
  \fmfpen{thin}
  \fmfleft{i}\fmfright{o}
  \fmf{plain}{i,v}\fmf{plain}{o,v}
  \fmf{dbl_plain}{v,v}
\end{fmfgraph}} +2\,
\parbox{15mm}{
\begin{fmfgraph}(40,20)
  \fmfpen{thin} \fmfleft{i}\fmfright{o}
  \fmf{plain}{i,v1}\fmf{plain}{o,v2}
  \fmf{dbl_plain,left,tension=0.3}{v1,v2}\fmf{dbl_plain,right,tension=0.3}{v1,v2}
\end{fmfgraph}} \nonumber \\&+& 18\,
\parbox{15mm}{
\begin{fmfgraph}(40,20)
  \fmfpen{thin}
  \fmfleft{i}\fmfright{o}
  \fmf{plain}{i,v1}\fmf{plain}{o,v2}
  \fmf{plain,left,tension=0.3}{v1,v2}\fmf{plain,right,tension=0.3}{v1,v2} \label{selfh}
\end{fmfgraph}}
\end{eqnarray}
The double lines indicate $\chi$ fields while single lines stand for $h$
ones. We are not considering tadpole diagrams which we require to
vanish at each given loop order by adding a counterterm linear in
$h$. The number in front of each diagram is the associated
combinatorial factor.

We consider the limit in which the $h$ mass is much larger than
the temperatures in play, hence all of the loops containing only
$h$ fields are then infrared finite at any temperature. The second
diagram in eq.~(\ref{selfh}) is, however, linearly ultraviolet
divergent but this divergence is absorbed in the respective mass
counterterm. After all the ultraviolet divergences have been taken
into account, we can show that the finite temperature corrections
due to $h$ are Boltzman suppressed. We are left with the following
infrared divergent graph for $T\rightarrow T_c$,
\begin{eqnarray}
\parbox{15mm}{
\begin{fmfgraph}(40,20)
  \fmfpen{thin} \fmfleft{i}\fmfright{o}
  \fmf{plain}{i,v1}\fmf{plain}{o,v2}
  \fmf{dbl_plain,left,tension=0.3}{v1,v2}\fmf{dbl_plain,right,tension=0.3}{v1,v2}
\end{fmfgraph}} = T\,\frac{g^2_1}{16\pi  m_{\chi}} \ , \label{one-loop}
\end{eqnarray}
and this diagram constitutes the relevant one loop correction to
the $h$ mass which reads:
\begin{eqnarray}
m^2(T)=m^2 -T\,\frac{g^2_1}{16\pi  m_{\chi}} \ .
\end{eqnarray}
This is the way the nearby phase transition is directly felt by
the non order parameter field. It shows that the mass of the
singlet field must decrease fast close to the phase transition. An
unpleasant consequence is that this one loop result breaks down at
the transition point due to the infrared singularity.

Besides, since $h$ is not the order parameter, its correlation
length (i.e. $1/m$) is not expected to diverge at the phase
transition. To cure the infrared behavior we need to go beyond the
one-loop approximation and consistently resum the following set of
diagrams:
\begin{eqnarray}
\parbox{15mm}{\begin{fmfgraph}(40,20)
    \fmfleft{i}\fmfright{o}
    \fmf{plain}{i,v1}\fmf{plain}{v2,o}
    \fmf{dbl_plain,left,tension=0.3}{v1,v2,v1}
\end{fmfgraph}} + \parbox{15mm}{
\begin{fmfgraph}(40,20)
  \fmfleft{i}\fmfright{o}
  \fmf{plain}{i,v1}\fmf{plain}{v3,o}
  \fmf{dbl_plain,left,tension=0.3}{v1,v2,v1}\fmf{dbl_plain,left,tension=0.3}{v2,v3,v2}
\end{fmfgraph}} + \parbox{15mm}{
\begin{fmfgraph}(40,20)
  \fmfleft{i}\fmfright{o}
  \fmf{plain}{i,v1}\fmf{plain}{v4,o}
  \fmf{dbl_plain,left,tension=0.3}{v1,v2,v1}
  \fmf{dbl_plain,left,tension=0.3}{v2,v3,v2}
  \fmf{dbl_plain,left,tension=0.3}{v3,v4,v3}
\end{fmfgraph}} + \cdot\cdot\cdot. \nonumber
\end{eqnarray}

Each diagram is infrared divergent but when resummed $m$ is finite
at $T$ and reads:
\begin{eqnarray}
m^2(T)=m^2 - T\, \frac{g_1^2}{16\pi\,m_{\chi}+\lambda\,T} \ .
\end{eqnarray}
At $T=T_c$ we have
\begin{eqnarray}
m^2(T_c)=m^2-\frac{g_1^2}{\lambda} \ .
\end{eqnarray}
Hence we predict that close to the phase transition the singlet
state has a decreasing mass. The drop at the phase transition
point is the ratio between the square of the coupling constant
governing the interaction of the singlet state with the order
parameter ($g_1$) and the order parameter self interaction
coupling constant ($\lambda$).

The slope is:
\begin{eqnarray}
{\cal D}^- \equiv\lim_{T\rightarrow T_c^-} \frac{1}{\Delta
m^2}\frac{d\, m^2(T)}{dT} =
\frac{16\,\pi}{\lambda\,T_c}\lim_{T\rightarrow T_c^-}\frac{d\,
m_{\chi}}{dT}\ ,
\end{eqnarray}
with $\Delta m^2= m^2(T_c) - m^2=g_1^2/\lambda$. This slope
encodes the critical behavior of the theory even though it is
constructed using the non order parameter field. For example if
$m^2_{\chi}$ vanishes as $(T_c-T)^{\nu}$ close to the phase
transition, then ${\cal D}^-$ scales with the exponent $(\nu/2
-1)$. When considering an $O(N)$ symmetry group rather than $Z_2$
the resummation procedure becomes exact in the large $N$ limit
\cite{Coleman:jh}.

When $T>T_c$ $\chi$ develops a vacuum expectation value which will
induce one also for $h$:
\begin{eqnarray}
\langle \chi^2 \rangle\equiv v^2=
3\,\frac{M^2_{\chi}}{\lambda-\frac{6\,g^2_1}{m^2}} \quad {\rm and}
\quad \langle h \rangle= -\frac{g_1}{2\,m^2} \langle \chi^2
\rangle \ ,
\end{eqnarray}
where we denoted the mass of $\chi$ in the broken phase by
$M_{\chi}^2=2\,|m_{\chi}|^2$ and considered small $h$ fluctuations
(i.e. we kept only the $g_1$ interaction term for $h$). This is in
agreement with the results found in \cite{Sannino:2002wb}. The
fields $\chi$ and $h$ mix in this phase and the mixing angle
$\theta$ is proportional to $g^2_1/m^2$. The
mixing can be 
neglected within the present approximations. Like for $T<T_c$, we
consider also now only the effects due to $\chi$ loops for the $h$
propagator. Due to symmetry breaking we now have the trilinear
$\chi$ couplings:
\begin{eqnarray} -\frac{\lambda}{3!}\,v\,\chi^3 \ ,\end{eqnarray}
which substantially affect the analysis and the results in this
phase. At the one loop level the diagram to compute is again the
one in eq.~(\ref{one-loop}) with $m_{\chi}$ replaced by
$M_{\chi}$. So we predict a drop on the right hand side of $T_c$
while this diagram is clearly infrared divergent. Curing the
divergence now is more involved due to the appearance of the
trilinear $\chi$ coupling. A typical diagram in the consistent set
of diagrams we consider is of the form:
\begin{equation}
\parbox{45mm}{
\begin{fmfgraph}(150,40)
    \fmfleft{i}\fmfright{o}
    \fmf{plain}{i,v1}\fmf{plain}{v9,o}
    \fmf{dbl_plain,left,tension=0.3}{v1,v2}
    \fmf{dbl_plain,left,tension=0.3}{v2,v1}
    \fmf{dbl_plain,left,tension=0.3}{v2,v3}
    \fmf{dbl_plain,left,tension=0.3}{v3,v2}
    \fmf{phantom,left,tension=0.3}{v3,v4}
    \fmf{phantom,left,tension=0.3}{v4,v3}
    \fmf{phantom,left,tension=0.3}{v4,v5}
    \fmf{phantom,left,tension=0.3}{v5,v4}
    \fmf{phantom,left,tension=0.3}{v5,v6}
    \fmf{phantom,left,tension=0.3}{v6,v5}
    \fmfv{decor.shape=circle,decor.filled=full,
          decor.size=0.5thick}{v4,v5}
    \fmf{dbl_plain,left,tension=0.3}{v6,v7}
    \fmf{dbl_plain,left,tension=0.3}{v7,v6}

    \fmf{phantom,left,tension=0.3,tag=1}{v7,v8}
    \fmf{phantom,left,tension=0.3,tag=2}{v8,v7}
    \fmf{phantom,left,tension=0.3}{v8,a1}
    \fmf{phantom,left,tension=0.3}{a1,v8}
    \fmf{phantom,left,tension=0.3}{a1,a2}
    \fmf{phantom,left,tension=0.3}{a2,a1}
    \fmf{phantom,left,tension=0.3}{a2,a3}
    \fmf{phantom,left,tension=0.3}{a3,a2}
    \fmfv{decor.shape=circle,decor.filled=full,
          decor.size=0.5thick}{a1,a2}
    \fmf{phantom,left,tension=0.3,tag=1}{a3,v9}
    \fmf{phantom,left,tension=0.3,tag=2}{v9,a3}

    \fmfposition
    \fmfipath{p[]}
    \fmfiset{p1}{vpath1(__v7,__v8)}
    \fmfiset{p2}{vpath2(__v8,__v7)}
    \fmfiset{p3}{vpath1(__a3,__v9)}
    \fmfiset{p4}{vpath2(__v9,__a3)}
    \fmfi{dbl_plain}{subpath (0,length(p1)/2) of p1}
    \fmfi{dbl_plain}{subpath (length(p1)/2,length(p1))of p1}
    \fmfi{dbl_plain}{subpath (0,length(p2)/2) of p2}
    \fmfi{dbl_plain}{subpath (length(p2)/2,length(p2)) of p2}
    \fmfi{dbl_plain}{point length(p1)/2 of p1 -- point length(p2)/2 of p2}
    \fmfi{dbl_plain}{subpath (0,length(p3)/2) of p3}
    \fmfi{dbl_plain}{subpath (length(p3)/2,length(p3))of p3}
    \fmfi{dbl_plain}{subpath (0,length(p4)/2) of p4}
    \fmfi{dbl_plain}{subpath (length(p4)/2,length(p4)) of p4}
    \fmfi{dbl_plain}{point length(p3)/2 of p3 -- point length(p4)/2 of p4}
\end{fmfgraph}}
\end{equation}
\end{fmffile}
We choose this subset of diagrams since the associate infinite sum
can be performed exactly \cite{AFK} and it has knowledge of the
onset of chiral symmetry breaking via the trilinear vertices.
\begin{eqnarray}
m^2(T)&=& m^2-\frac{g^2_1{\cal
I}}{2}\,\frac{1+\frac{\lambda}{3}{\cal
I}}{1+\frac{\lambda}{2}{\cal I}+ \frac{\lambda^2}{6}{\cal I}^2} \
, \\{\cal I}&=&\frac{1}{8\pi\,M_{\chi}}\ .
\end{eqnarray}
The infrared divergence has been cured for $T\geq T_c$. We also
see that $m(T_c)$ from the right hand side of the transition
equals exactly the one from the unbroken side of the transition.
This remarkable result does not hold order by order in the loop
expansion but only when an infinite sum of the diagrams is
performed. The $h$ mass square is a continuous function throughout
the phase transition and the associated correlation length remains
finite. However the slope of the $h$ mass near the phase
transition from the right hand side does not mirror the one on the
left. This fact is due to the onset of spontaneous symmetry
breaking communicated to $h$ via the trilinear $\chi$ interaction
term. Indeed:
\begin{eqnarray} {\cal D}^+
&\equiv&\lim_{T\rightarrow T_c^+} \frac{1}{\Delta
m^2}\frac{d\,m^2(T)}{dT}\nonumber \\
&=&\frac{3}{2}\left(\frac{16\pi}{\lambda T_c}\right)^2\,
\lim_{T\rightarrow T_c^+}\frac{dM^2_{\chi}}{dT}\nonumber
\\&=&{3}\left(\frac{16\pi}{\lambda T_c}\right)^2\, \lim_{T\rightarrow
T_c^+}\frac{d|m_{\chi}|^2}{dT} \ ,
\end{eqnarray}
and we deduce the relation:
\begin{eqnarray}
{\cal D}^+ =
-6\,\frac{16\pi}{\lambda}\frac{|m_{\chi}|}{T_c}\,{\cal D}^- \ .
\end{eqnarray}
The $h$ mass drops on the left hand side of the phase transition
and rises on the right one. The slope on the right side (broken
phase) is less singular with respect to the one on the left. More
specifically the scaling exponent for ${\cal D}^+$ is $(\nu -1)$.
Deviations from such a behavior directly measure departures from
the second order character of the phase transition. In the figure
we schematically represent the behavior of the $h$ mass as a
function of the temperature in units of the critical temperature
for $m_{\chi}^2\propto \left(T_c-T \right)$.
\begin{figure}[]
 \includegraphics[width=8.8 truecm, height=3truecm]{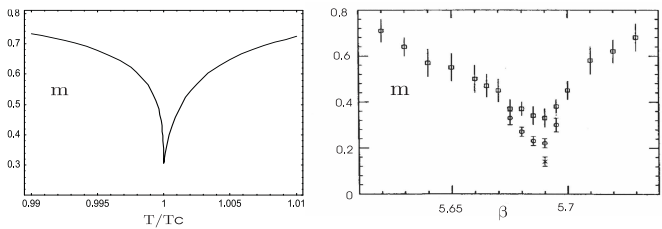}
\caption{Left panel: behavior of the mass of the singlet field
close to the phase transition as function of the temperature.
Right panel: Lattice data; figure from \cite{Bacilieri:mj}.}
\label{Figura1}
\end{figure}
The similarity of our result with the lattice data
\cite{Bacilieri:mj} (right panel of Fig.\ref{Figura1}) is
surprising. Recall, though, that our results are directly
applicable to the two color Yang-Mills theory, while the data in
\cite{Bacilieri:mj} are obtained for the three color Yang-Mills
theory. This resemblance is in agreement with the almost second
order character of the three color deconfining phase transition.
Another difference is that the mass in \cite{Bacilieri:mj} seems
to be the screening one while we are plotting here the physical
mass.

It is of common practice to isolate the order parameter and more
generally the light degrees of freedom since they are expected to
be the relevant degrees at low energies. While this is certainly
correct we have explicitly shown how all of the essential features
of the phase transition are encoded in the non order parameter
fields of the theory as well. Our analysis is  useful whenever the
order parameter is neither a physical quantity nor
phenomenologically accessible (e.g. the Polyakov Loop).

We also studied what happens if the order parameter depends on the
time coordinate. The $h$ mass has the same behavior near the
transition as in the previous case. This can be easily understood
since the infrared physics is dominated by the zero mode of
$\chi$. The major difference with the previous analysis is that a
physical width develops describing the decay of $h$ into two
$\chi$ fields at non zero temperature \cite{AFK}.

\section{Deconfinement and Conclusions}
\label{due}

The $SU(N)$ deconfinement phase transition is a hard problem.
Importance sampling lattice simulations provide crucial
information about the nature of the temperature driven phase
transition for 2 and 3 colors Yang-Mills theories with and without
matter fields (see \cite{Boyd:1996bx} for 3 colors). Different
approaches
\cite{Sannino:2002wb,Agasian:fn,Campbell:ak,Simonov:bc,Sollfrank:du,Carter:1998ti,Pisarski:2001pe,DP,KorthalsAltes:1999cp,Dumitru:2001xa,Wirstam:2001ka,Laine:1999hh,Sannino:2002re}
are used in literature to tackle the features of this phase
transition. At zero temperature $SU(N)$ Yang-Mills theory is
asymptotically free and the physical spectrum of the theory
consists of a tower of hadronic states referred to as glueballs
and pseudo-scalar glueballs. The theory develops a mass gap and
the lightest glueball has a mass of the order of few times the
confining scale.

At nonzero temperature the $Z_N$ center of $SU(N)$ is a relevant
global symmetry \cite{Svetitsky:1982gs}, and it is possible to
construct a number of gauge invariant operators charged under
$Z_N$ among which the most notable one is the Polyakov loop:
\begin{eqnarray} {\ell}\left(x\right)=\frac{1}{N}{\rm Tr}[{\bf L}]\equiv\frac{1}{N}{\rm Tr}
\left[{\cal
P}\exp\left[i\,g\int_{0}^{1/T}A_{0}(x,\tau)d\tau\right]\right] .
\end{eqnarray} ${\cal P}$ denotes path ordering, $g$ is the $SU(N)$ coupling constant and $x$ is
the coordinate for the three spatial dimensions while $\tau$ is
the euclidean time. The $\ell$ field is real for $N=2$ while
otherwise complex.  This object is charged with respect to the
center $Z_N$ of the $SU(N)$ gauge group \cite{Svetitsky:1982gs}
under which it transforms as $\ell \rightarrow z \ell$ with $z\in
Z_N$. A relevant feature of the Polyakov loop is that its
expectation value vanishes in the low temperature regime while is
non zero in the high temperature phase. The Polyakov loop is a
suitable order parameter for the Yang-Mills temperature driven
phase transition \cite{Svetitsky:1982gs}.

Pisarski used this feature \cite{Pisarski:2001pe} to model the
Yang-Mills phase transition via a mean field theory of Polyakov
loops. The model is referred to as the Polyakov Loop Model (PLM).
Recently some interesting phenomenological PLM inspired models
aimed to understand RHIC physics were constructed
\cite{Scavenius:2001pa,{Scavenius:2002ru}}.

Here we consider pure gluon dynamics. This allows us to have a
well defined framework where the $Z_N$ symmetry is exact. The
hadronic states of the Yang-Mills theory are the glueballs. At
zero temperature the Yang-Mills trace anomaly has been used to
constrain the potential of the lightest glueball state $H$
\cite{Schechter:2001ts}:
\begin{eqnarray}
V[H]=\frac{H}{2} \ln \left[\frac{H}{\Lambda^4}\right] .
\end{eqnarray}
$\Lambda$ is chosen to be the confining scale of the theory and
$H$ is a mass dimension four field. This potential correctly
saturates the trace anomaly when $H$ is assumed to be proportional
to ${\rm Tr}\left[G_{\mu \nu}G^{\mu \nu}\right]$ and $G_{\mu \nu}$
is the standard Yang-Mills field strength.

In \cite{Sannino:2002wb} a concrete model was proposed able to
transfer the information about the Yang-Mills phase transition
encoded in the $Z_N$ global symmetry to the hadronic states of the
theory. This model is constructed using trace anomaly and the
$Z_N$ symmetry. The model in \cite{Sannino:2002wb} is supported by
recent theoretical investigations \cite{Meisinger:2002ji}. Once
the relation between the fields $H$ and $h$, and $\ell$ and $\chi$
is made the relations between the couplings of the Lagrangian for
$H$ and $\ell$ constructed in \cite{Sannino:2002wb} and the
renormalizable one presented here can be obtained. More
specifically, our renormalizable theory is a truncated (up to
fourth order in the fields) version of the full glueball theory.
For example one can take the following relation between $H$ and
the glueball field $h$:
\begin{eqnarray} H=\langle H \rangle \left( 1+
\frac{h}{\sqrt{c}\langle H \rangle^{1/4}}\right) \ .\end{eqnarray}
Here $\langle H\rangle=\Lambda^4/e$ is the vacuum expectation
value of the glueball field below the critical temperature and $c$
is a positive dimensionless constant fixed by the mass of the
glueball. For the $\ell$ field we have $\chi=\sqrt{\kappa}\ell $
with  $\kappa$ a mass dimension two constant (at high temperature
is proportional to $T^2$).

Our present results are simply the higher loop corrections to the
glueball model presented in \cite{Sannino:2002wb} and can be
immediately applied to the two color Yang-Mills phase transition.
We confronted already our theoretical results with lattice
computations of the glueball mass behavior close to the phase
transition studied in \cite{Bacilieri:mj} for three colors. Our
analysis not only is in agreement with the numerical analysis but
allows us to provide a better understanding of the physics in
play. More recent lattice results should provide a still more
detailed test of our predictions.

Finally, using a renormalizable theory containing a singlet heavy
field uncharged under a global symmetry and the associated order
parameter with their possible interactions we demonstrated that
the information about the phase transition can be extracted via
the singlet field. The transfer of information from the order
parameter to the heavy field is in fact complete. Moreover, if we
consider the large $N$ limit for the $O(N)$ theory the analysis in
the unbroken phase becomes exact. Thanks to the result established
here and envisioned in \cite{Sannino:2002wb} we can use directly
the knowledge about $h$, or more specifically of some of its
properties defined in the text, to characterize the phase
transition. We considered a $Z_2$ symmetry group appropriate for
example to discuss the deconfining phase transition for 2 color
Yang-Mills theory. The generalization to a general group of
symmetries is straightforward.

\acknowledgments It is a pleasure to thank P.H. Damgaard for
enlightening discussions and careful reading of the manuscript. We
thank J.~Schechter for valuable discussions and careful reading of
the manuscript and K. Rummukainen for encouragement and
discussions. The work of F.S. is supported by the Marie--Curie
fellowship under contract MCFI-2001-00181.

\end{document}